\documentstyle[preprint,aps,epsfig]{revtex}
 
\newcommand{\be}{\begin{equation}}
\newcommand{\ee}{\end{equation}}
\newcommand{\bea}{\begin{eqnarray}}
\newcommand{\eea}{\end{eqnarray}}
\newcommand{\lessim}{\stackrel{<}{\sim}}
\newcommand{\bm}{\boldmath}
\newcommand{\fr}[2]{\frac{{\displaystyle #1}}{{\displaystyle #2}}}
\newcommand{\ggam}{\mbox{$\gamma\gamma\,$}}
\tighten
\begin{document}
\draft
\preprint{UL--NTZ 02/98}
\title {
    Search for a heavy magnetic monopole \\ at the Fermilab 
    Tevatron and CERN LHC}
\author{
   I.F. Ginzburg\thanks{ginzburg@math.nsc.ru}}
\address{    
         Institute of Mathematics,  630090 Novosibirsk, Russia}
\author{
   A. Schiller\thanks{schiller@tph204.physik.uni-leipzig.de}}
\address{   
    Institut f\"ur Theoretische Physik and NTZ, \\
    Universit\"at Leipzig, D-04109 Leipzig, Germany}
\date{Revised March 16, 1998}
\maketitle

\begin{abstract}
If a heavy Dirac monopole exists, the light to light scattering
below the monopole production threshold is enhanced due to the
strong coupling of monopoles to photons. This effect could be
observable in the collision of virtual photons at proton
colliders. At the Tevatron it will be seen as pair production of
photons with energies 200--400 GeV and roughly compensated
transverse momenta 100--400 GeV/c.  This effect could
be seen at monopole masses about 1--2.5 TeV at the upgraded Tevatron
and 7.4--19 TeV at LHC depending on monopole spin.
\end{abstract}


\narrowtext

\section{Introduction}

The magnetic charge (monopole) was introduced into particle
theory by Dirac \cite{1} (see also \cite{2}) to restore symmetry
among electricity and magnetism.  The idea of a monopole is very
attractive in order to explain the mysterious quantization of
the electric charge of particles. Therefore, the search for a
monopole is essential despite the fact that there is no place
for it in the standard description of our world.

The Dirac--Schwinger monopoles, discussed here, are {\em
point--like} particles. They differ strongly from non--local
monopoles in the context of gauge theories, first considered by
Polyakov and 't Hooft \cite{gaugemono}. Below we assume that a
few monopoles exist in the Universe and they were not yet
observed due to their high mass.

Our basic idea is simple: {\em The existence of monopoles
provides for a $\gamma\gamma$ elastic scattering at large
angles, that is sufficiently strong below the monopole
production threshold. This effect is observable at colliders
with energies smaller than monopole masses.} Such a method for
discovering monopoles has been suggested in Refs.
\cite{GPanf,BBG}. A similar idea for the process $e^+e^-\to Z\to
3\gamma$ has been considered recently \cite{Ruj,BBCL} and tested
at LEP \cite{Acc}. Our paper deals with detailed calculations of
this effect at hadron colliders.  Throughout the paper we denote
the monopole mass by $M$ and its spin by $J_M$ (we assume a
definite spin of the monopole), $\omega$ is the characteristic
photon energy (typically -- the $\ggam$ c.m.s. energy).

A theory with two point--like charges, electric and magnetic,
cannot be standard QED\footnote{According to \cite{he}, this
theory even could violate $U(1)$ local gauge invariance of
standard QED. We are thankful to Dr. He for clarification of this
point.}. According to Refs. \cite{1,2} (see also \cite{Ruj}),
the electromagnetic field in such a theory is described by a
vector potential having the Dirac string or some of its
surrogate. To have unambiguous results, the elementary electric
and magnetic charges $e$ and $g$ ought to be quantified so
that\footnote{In the Schwinger theory \cite{2} the quantity
$n$ should be even, $n=\pm 2, \pm 4,\dots$}
\be
  g=\fr{2\pi n}{e}\,, \quad n= \pm 1,\,\pm 2,\dots
  \label{dirac}
\ee
with $\alpha\equiv e^2/(4\pi)=1/137 \Rightarrow\alpha_g
\equiv g^2/(4\pi) =n^2/(4\alpha)$. Unfortunately, the explicit
form of such a theory is still not yet known.

We are interested in the energy region below the monopole
production threshold. Here the interactions of photons via
virtual monopoles are considered as main effect. 
The effective expansion parameter is of the order of
\be
  g_{\mathrm{eff}}= \fr{g\omega}{\sqrt{4\pi}M}\equiv\fr{n\omega}{
  2\sqrt{\alpha} M}< 1.
\ee
In this region general considerations like gauge invariance, threshold
behavior, etc. together with a perturbative approach allow us to
believe that a perturbation theory analogous to standard QED can
be applied (certainly, in lowest nontrivial order only).  In
that case QED--like calculations should be valid, both in tree
approximation and at one loop level.  For one loop we can assume
a Wick rotation into Euclidean region. As it is well known, the
integration over the loop momentum $p$ 
is convergent due to gauge invariance. Therefore,
the integration region is limited by virtualities
$p^2\lessim\omega^2$, where the effective expansion parameter is
small, and QED results are valid.

\begin{figure}[!htb]
     \begin{center}
      \leavevmode
      \epsfig{file=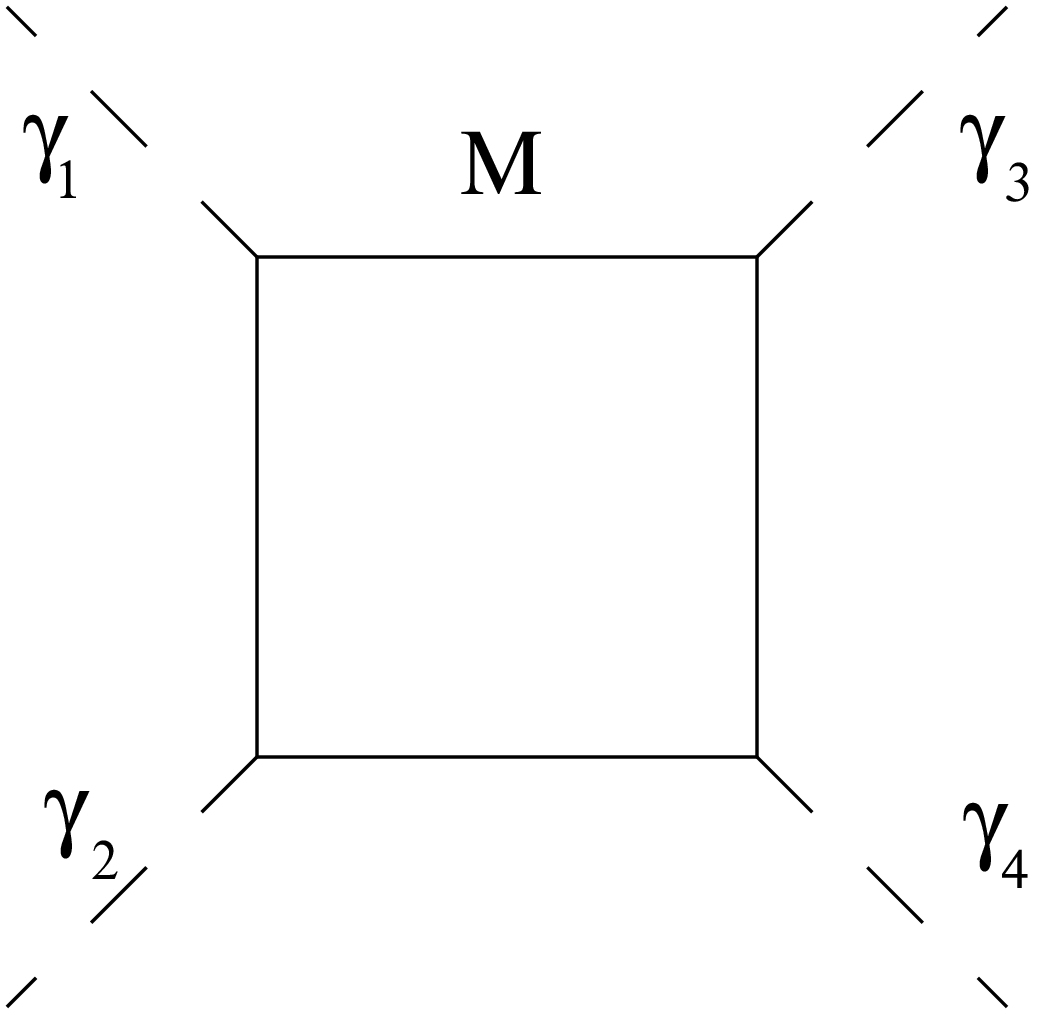,height=65mm,width=65mm,angle=270}
      \caption{$\ggam \to \ggam$ via monopole loop}
      \label{fig:1}
    \end{center}
\end{figure}

The estimate of the effective coupling constant is supported 
by estimating qualitatively
the cross section of $\ggam\to\ggam$ scattering via a monopole
loop (Fig.\ \ref{fig:1}) 
in the \ggam\ c.m.s. (without fixing the form of the interaction).
Due to gauge invariance, each photon
leg of the matrix element yields a factor $\omega$.  On
dimensional grounds this factor has to be divided by the
monopole mass $M$. Additionally, the magnetic charge $g$ has to
be associated to each vertex. Therefore, the amplitude ${\cal
M}\propto\left({g\omega}/{M}\right)^4$ and the cross section 
is\footnote{ If some monopole --
antimonopole bound state with mass $\ll 2M$ exists, it is wide
enough (since $g\gg 1$), and it gives additional more strong
light to light scattering. We don't speculate about this
opportunity in detail.} 
$\sigma\propto ({1}/{\omega^2})\left({g\omega}/{M}\right)^8$.

The corresponding effective Lagrangian (of Heisenberg--Euler
type) is
\be
  {\cal L}_{\mathrm{eff}} = -\fr{F_{\mu \nu}F^{\mu \nu}}{4}
  +\fr{g^4}{36(4\pi)^2\,
  M^4}\left[\fr{\beta_++\beta_-}{2}
  \left(F^{\mu\nu}F_{\mu\nu}\right)^2
  +\fr{\beta_+-\beta_-}{2}
  \left(F^{\mu\nu}\tilde{F}_{\mu\nu}\right)^2\right]
  + \ldots
\ee
with the electromagnetic field strength tensor $F_{\mu\nu}$ and
$\tilde{F}^{\mu \nu}= \epsilon^{\mu \nu \alpha \beta}
F_{\alpha\beta}/2$. An additional gauge fixing term has to be
introduced to invert the photon propagator. The constants
$\beta_{\pm}$ depend on the monopole spin (a numerical
coefficient is introduced to simplify the final expressions).
Based on arguments mentioned earlier (related to Wick rotation)
we use here the coefficients $\beta_\pm$ obtained in QED.
Their values have been found for different values of the monopole spin
$J_M$ in \cite{HE} ($J_M=1/2$), \cite{spin0} ($J_M= 0$),
\cite{BB,Jik} $(J_M=1)$. We use their combination $P =
\beta_+^2 +2\beta_-^2$. 

Let us denote the standard Mandelstam variables for the \ggam\
scattering by $\hat{s}$, $\hat{t}$, $\hat{u}$ with $\hat{s}+
\hat{t}+\hat{u}=0$ ($\hat{s}$ is the effective mass of
produced \ggam\ system squared and $\hat t$ varies within the
interval ( $0,\; -\hat{s}/2$)). Furthermore, $\theta$ is the
scattering angle in c.m.s. of the photons with $0<\theta
<{\pi}/{2}$. In our case $\hat{s}= 4\omega^2$, $\hat{t}=
-2\omega^2(1-\cos\theta)$.

The cross section for $\ggam$ scattering via a monopole loop at
$\hat{s} \ll M^2$ is given by the expressions
\be
  \sigma^{\mathrm{tot}} =
  R\left(\fr{\hat{s}}{4}\right)^3\,,\quad
  R= \fr{28P}{405\pi}\left(\fr{n}{2\sqrt{\alpha}M}\right)^8\,,
  \label{sigtot}
\ee
\bea
  d\sigma&=&
  \fr{5(3+\cos^2\theta)^2}{56}\sigma^{\mathrm{tot}}d\cos\theta\equiv
  \fr{5}{7}\left(\fr{\hat{s}^2 +\hat{t}^2
  +\hat{u}^2}{\hat{s}^2}\right)^2 \sigma^{\mathrm{tot}}
  \fr{d\hat{t}}{\hat{s}}\,,
  \label{dsiggg} 
\eea
\begin{equation}
  P=\beta_+^2 +2\beta_-^2 =\left\{
  \begin{array}{ll} 
    0.085,  &  J_M=0   \\
    1.39,   &  J_M=1/2 \\
    159 ,   &   J_M=1 
  \end{array} 
  \right..
\end{equation}
Note that the result strongly depends on
the monopole spin $J_M$.

These expressions are correct up to contributions of order
${\cal{O}}(g^2_{\mathrm{eff}})$. We expect that with growing
$g_{\mathrm{eff}}$ the increase of the cross section as a
function of energy stops and the production of a larger (even)
number of photons becomes essential. The effective parameter
here is expected to be less than $g_{\mathrm{eff}}^2$. For
example, in QED with spin 1/2 fermions, the ratio of
coefficients of the $6\gamma$ to $4\gamma$ operators in the
effective Lagrangian for $\ggam$ interactions is about $0.3
g_{\mathrm{eff}}^2$ \cite{HE}.  Besides, one can expect that
below the threshold ($\omega<M$) the expressions
(\ref{sigtot}) correctly describe the sum of cross sections
of processes $\ggam\to 2\gamma,\;4\gamma,\;6\gamma,\ldots$ with
increasing multiple photon production at higher
$g_{\mathrm{eff}}$.

\section{{\bm $\ggam$} production via a monopole loop at hadron colliders}
\label{s:2}

The cross section for the production of two photons via a
virtual heavy monopole loop in $pp$ or $p\bar{p}$ collisions is
a convolution of the $\ggam$ cross section (\ref{dsiggg}) with the
photon fluxes arising from the colliding protons
\be
  \sigma_{pp\to \gamma\gamma X}=\left(\fr{\alpha}{\pi}\right)^2
  \int n_1(\omega_1,Q_1^2) n_2(\omega_2,Q_2^2) \fr{d\omega_1}{\omega_1}
  \fr{d\omega_2}{\omega_2}\fr{dQ_1^2}{Q_1^2}
  \fr{dQ_2^2}{Q_2^2}d\sigma_{\gamma\gamma\to
  \gamma\gamma}.\label{prottot}
\ee
Since in the dominant integration region $Q_i^2 \ll\hat{s}$, the
transverse motion of initial photons can be neglected with good
accuracy.

The relevant scale for the virtuality dependence of the $\ggam
\to\ggam$ subprocess cross section is given by the unique
inner parameter of monopole loop -- the monopole mass.
Therefore, this dependence appears in the form of the quantity $Q^2/M^2$. 
Since $Q^2\ll \hat{s} \ll M^2$, it is safely neglected below.

Taking these approximations into account the cross section
(\ref{prottot}) can be written in factorized form (valid for
$Q_i^2 \ll \hat{s}$). The photon flux densities $n_i$ depend on
their ``own'' photon variables only
\be
  n_i\equiv n_i(y_i=\fr{\omega_i}{E},Q_i^2)\,,\quad  
  \hat{s}=4\omega_1\omega_2
\ee
and we have  for the cross section
\bea
  \sigma_{pp\to\gamma \gamma X}& =&
  \left(\fr{\alpha}{\pi}\right)^2 \int \fr{dy_1}{y_1} f(y_1)
                                       \fr{dy_2}{y_2} f(y_2)
  d\sigma_{\gamma\gamma\to \gamma\gamma}
  \equiv
  \left(\fr{\alpha}{\pi}\right)^2 R\, E^6
  N^2(E)\,, \label{crsec}\\
  &&N(E)=\int y^2 f(y)dy\,,\quad   f(y) = \int n(y,Q^2)
  \fr{dQ^2}{Q^2}\,.\label{ne}
\eea

The photon flux density arising from one proton is a 
sum over elastic and inelastic contributions:
\begin{equation}
  n(y,Q^2)=(D_{\mathrm{el}}+D_{\mathrm{in}})
   + \fr {y^2}{2}(C_{\mathrm{el}}+C_{\mathrm{in}})\,.
  \label{ny}
\end{equation}
In the individual contributions the quantity $Q^2$ is limited
kinematically from below ($m_p$ is the proton mass):
\be
  Q^2>Q^2_{\mathrm{min}} = (M_X^2-m_p^2)\fr{y}{1-y}
  +m_p^2\fr{y^2}{1-y}\,.\label{min}
\ee
$M_X$ is the effective mass of the system produced
in the virtual $\gamma^*p$ collision,
$M_X=m_p$ in the elastic case.

{\em The elastic contribution} is written via standard proton
form factors $G_E$ and $G_M$
\bea
  C_{\mathrm{el}}(Q^2)&=& G_M^2(Q^2)\,,
  \nonumber \\
  D_{\mathrm{el}}(Q^2)&=& (1-y)\fr{ 4 m_p^2 G_E^2(Q^2)
  +Q^2 G_M^2(Q^2) }{ 4 m_p^2
  +Q^2}\left(1-\fr{Q^2_{\mathrm{min}}}{Q^2}\right)\,.
  \label{elast}
\eea
The integral over $Q^2$ in (\ref{ne}) rapidly converges at the
upper boundary due to the form factors. The integral is
saturated at $Q^2$ values given by  
the form factor scale. This scale is much lower than the other
parameters of the problem. Therefore, the upper integration
limit can be extended to infinity and the elastic contribution
to $N(E)$ becomes energy independent.

{\em The inelastic contribution} is written via the proton
structure functions $F_1\,,F_2$. $C_{\mathrm{in}}$ and
$D_{\mathrm{in}}$ are integrals over the squared effective mass
of the produced system $M_X^2$.

We basically start from  Eq.\ (D.4) and Table~8 of Ref. \cite{9}
\bea
  C_{\mathrm{in}}(Q^2)&=&
  \fr{2}{Q^2} \int d M_X^2 \, F_1(M_X^2,Q^2) \,, \nonumber\\
  D_{\mathrm{in}}(Q^2)&=& \fr{(1-y)}{ Q^2} 
  {\int d M_X^2 \, x F_2(M_X^2,Q^2)
  \left( 1- \fr{Q^2_{\mathrm {min}}}{Q^2}\right)}\,.
\eea
In order to use the standard representations of the structure
functions we change the integration variable from $M_X^2$ to the
Bjorken variable $x$ using the relation $M_X^2=m_p^2+ Q^2
(1-x)/x$.  
Next, from inequality (\ref{min}) the lower limit in $x$ is found
\be
   x_{\mathrm{min}}=  \fr{y}{1-y^2 m_p^2/Q^2} \,,
   \label{xmin}
\ee
yielding the useful relation
\begin{displaymath}
  (1-y) x ( 1- \fr{Q^2_{\mathrm min}}{Q^2}) =
  y \left(\fr{ x}{x_{\mathrm {min}}}- 1 \right)\,.
\end{displaymath}
The upper $x$ value is reached at the minimal value for the
quantity $ L^2=M_X^2-m_p^2\approx2m_pm_\pi$:
\be
  x_{\mathrm{max}}=\fr{ Q^2}{Q^2+L^2} \,.
\ee
With these transformations we present $C_{\mathrm{in}}(Q^2)$ and
$D_{\mathrm{in}}(Q^2)$ in the form
\be
  C_{\mathrm{in}}(Q^2)= 2 \int\limits_{x_{\mathrm
  {min}}}^{x_{\mathrm {max}}} \frac{d x} {x^2} F_1(x,Q^2) \,,\quad
  D_{\mathrm{in}}(Q^2)= y \int\limits_{x_{\mathrm
  {min}}}^{x_{\mathrm {max}}} \fr{d x} {x^2}
  \left(\fr{x}{x_{\mathrm {min}}}- 1 \right) F_2(x,Q^2)\,.
\ee

Concerning the upper $Q^2$ limit for the inelastic contribution,
one should keep in mind that the basic representation with photon flux
factorization is valid at $Q_i^2 \ll \hat{s}$ only.  Near this
bound the original integrand is less than that obtained using
our factorization, but its contribution to the total cross
section is small. Therefore, the integration region can be
restricted from above by $Q_i^2 \ll \hat{s}=y_1y_2s$ with
sufficient accuracy. Since the virtual photon energies in
both fluxes are roughly equal, we use
\be
  Q_{i\,\mathrm{max}}^2 \approx y_i^2 s
\ee
as upper limit for the inelastic contribution without further
reducing the accuracy. With this choice the factorization
remains valid.

In the $y$ integration of (\ref{ne}) $y_{\mathrm{max}}=1$ can be
used as upper bound. The inaccuracy in the quantities that enter
(structure functions) is inessential since contributions at
$y_i\approx 1$ can be safely neglected.

\section{Total cross section and mass limits}

{\em Elastic contribution.} 
The proton form factors are
written in the standard dipole approximation
\be
  \fr{ G_M(Q^2)}{\mu_p} =  G_E(Q^2) =
   \frac{1}{\left(1+{Q^2}/{Q_0^2}\right)^2}\,,\quad
  \mu_p=2.79285\,,\quad Q_0^2=
  0.71\,\mathrm{GeV}^2\,.
   \label{form}
\ee
Using dimensionless variables $ a= {4 m_p^2}/{Q_0^2}$ and
$z=({a}/{4}) \times {y^2}/({1-y})$ the energy distribution is
found\footnote{ Compare \cite{9}, Eq.\ (D.7) with a sign misprint
corrected and upper $Q^2$ limit neglected.} as follows
\bea
  \fr{dN_{\mathrm{el}}}{d y}& = &\fr{y^2 (1-y)}{a} \Big[
  \left(a + z \left(1 +\mu_p^2 + 4 a \right) \right) I(z,0) \Big.
  \nonumber \\
  &&\Big. + (a+z) (\mu_p^2-1) I(z,a) -  \fr{a}{ (1+z)^3}
   \Big]
\eea
where
\begin{displaymath}
   I(z,a)=- \fr{1}{(1-a)^4} \left( \log \fr{a+z}{1+z} +
   \sum_{k=1}^{3} \fr{1}{k}\left( \fr{1-a}{1+z}\right)^k  \right)\,.
\end{displaymath}
With the numerical values for $\mu_p$ and $Q_0$ the remaining
$y$ integration leads to the energy independent constant
$N_{\mathrm{el}}=0.017672$ as elastic contribution to $N(E)$.

{\em Inelastic contribution.} The structure functions
$F_{1,2}$ are parametrized using the next to leading order quark
distributions of \cite{grv} and the parton model relation $F_2=
2 x F_1$. To test the sensitivity of our results on this
particular parametrization, we have repeated some of the
calculations using the structure functions of \cite{mrrs}. The
difference is small enough. 
 
The $Q^2$ dependence of the $F_{1,2}$ parametrizations 
is valid above some low input value $Q^2_{\mathrm{low}}$. 
Fortunately, the inelastic contributions
to $N(E)$ 
should be almost insensitive to the behavior of structure
functions at small $Q^2$. To test this statement, we consider
two extrapolations for these functions below
$Q^2_{\mathrm{low}}$: $F_{1,2} (x,Q^2) =0$ or $F_{1,2}(x,Q^2)
=F_{1,2}(x,Q^2_{\mathrm {low}})$.  At $E=0.9$ TeV in GRV
parametrization \cite{grv} the results coincide within $1 \%$
accuracy.

{\em Numerical estimates.} At 0.9
TeV we obtain $N(E)=0.0410$ in 
the GRV and $N(E)=0.0338$ in the MRRS approximation \cite{mrrs}
for this factor of Eq.\ (\ref{crsec}). The quantity $N(E)$
depends only weakly on the proton energy $E$.  The ratio
$N(E)/N(0.9\mbox{ TeV})$ varies from 0.966 at $E=0.5$ TeV to
1.006 at 1 TeV and 1.102 at $E=7$ TeV using GRV parametrization.

For a monopole with $J_M=1/2$ and $M/n=1$ TeV and a proton
energy of 1 TeV we obtain the total cross section
\be
  \sigma_{pp\to \gamma \gamma X}=150 \, {\mathrm {fb}}\,.
  \label{total}
\ee
It is useful to rewrite this photon production cross section for
different proton collider energies and different kinds of
monopoles
\be
  \sigma_{pp \to \gamma\gamma X}(E,M,P,n)=
   108 \, P \, \left(\fr{n E}{M}\right)^8
   \left( \fr{N(E) }{N(1 \,{\mathrm {TeV}}) } \right)^2
   \left( \fr{ 1\,{\mathrm {TeV}}}{  E } \right)^2
     {\mathrm {fb}}\,.
\ee

Let us consider a luminosity integral of 2 fb$^{-1}$ and a beam
energy of 0.9 TeV (Tevatron). If we assume 10 events to be
sufficient to detect the discussed effect,
the following mass limits can be reached (for different
spins $J_M$)
\begin{equation}
  M<n\otimes\left\{\begin{array}{rcl}
  0.998 &{\text{TeV}}\,,&J_M=0  \\
  1.42  &{\text{TeV}}\,,&J_M=1/2 \\
  2.56  &{\text{TeV}}\,,&J_M=1
\end{array}\right..
\end{equation}
Taking 100 fb$^{-1}$ and $E=7$ TeV (LHC) we obtain the following
mass limits
\begin{equation}
  M<n\otimes\left\{\begin{array}{rcl}
  7.40 &{\text{TeV}}\,, &J_M=0   \\
  10.5 &{\text{TeV}}\,, &J_M=1/2 \\
  19.0 &{\text{TeV}}\,, &J_M=1
\end{array}\right..
\end{equation}

The obtained limiting quantities correspond to a $\ggam\to\ggam$
subprocess cross section (calculated as $R\;\langle\omega\rangle^6
$  with an estimate of $\langle\omega\rangle$  taken from
(\ref{aomega}) below) which is roughly 500 pb for the Tevatron and
10 pb for LHC independent on monopole spin and $n$. These values
are much higher than the cross section of the main background
process $\ggam\to \ggam$ via a $W$ boson and a $t$ quark loop which
is about 20-30 fb \cite{Jik}.

\section{Energy and momentum distributions}

{\em Energy distribution for virtual photons.}
The photon fluxes in Eq.\ (\ref{crsec}) decrease with increasing
photon energies. On the other hand, the $\ggam$ subprocess cross
section rapidly increases with $\hat{s}= 4\omega_1\omega_2$.
Therefore, the main contribution to the cross section is given
by region of intermediate $\omega_i$. As already mentioned,
the dependence of the subprocess cross section on $Q^2$ can be neglected
since the characteristic values of virtuality $Q^2\ll\hat{s}\ll M^2$.
Therefore, the energy distribution for virtual photons is given
by (compare Eqs.\ (\ref{crsec},\ref{ne}))
\be
  \fr{d^2 \sigma}{dy_1 dy_2} = \left(\fr{\alpha} {\pi} \right)^2 R \, E^6
  \; y_1^2f(y_1) \; y_2^2 f(y_2) \,.
\ee

\begin{figure}[!htb]
     \begin{center}
      \leavevmode
      \epsfig{file=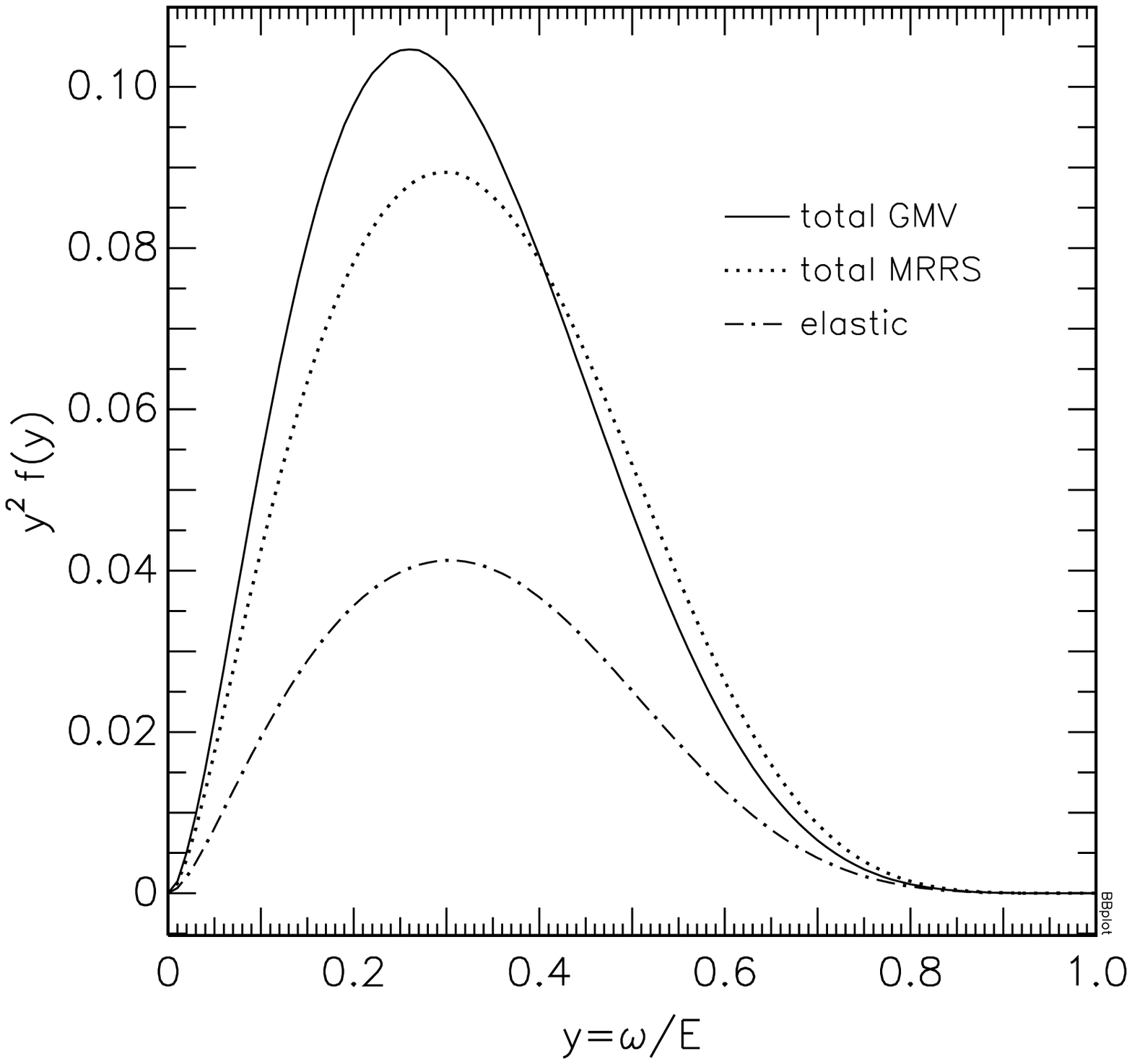,height=65mm,width=65mm}
            \caption{Energy distribution of virtual photons
           from the proton, $y^2 f(y)$ at 0.9 TeV}
      \label{fig:2}
    \end{center}
\end{figure}

Fig.\ \ref{fig:2} shows the energy distribution $y^2 f(y)$ for
photons arising from one proton (the distribution is identical
for each photon) using the two structure function
parametrizations.
The virtual photon energy distribution varies only weakly with
the energy $E$, this weak energy dependence manifests itself
in  the weak $E$ dependence of $N(E)$
mentioned above.  At 0.9 TeV the average energy of the colliding
photons and their energy spread are
\be
      \langle \omega \rangle =0.314 \, E\,,\quad
      \langle \Delta \omega \rangle =0.149 \, E \,.
\label{aomega}
\ee

{\em $\ggam$ differential cross sections.} 
As noticed at
the beginning of Sec.\ \ref{s:2}, the transverse motion of
incident (virtual) photons can be neglected with reasonable
accuracy.  Therefore, the transverse momenta of the
produced photons are balanced: $p_{T3}\approx -p_{T4} \equiv
p_T$.  The 4--momenta of these photons in the c.m.s.  of the
protons can be written in two forms
\be
    p_{3,4}=(\varepsilon_{3,4},\pm p_T,0,p_{L3,L4})\equiv
    p_T(\cosh\eta_{3,4},\pm 1,0,\sinh\eta_{3,4}) \,.   \label{kinem}
\ee
Here transverse and longitudinal momenta and rapidities of
photons are introduced.  Using these notations we have
\be
  \sin\theta=\fr{p_T}{\sqrt{\omega_1\omega_2}}\,,
  \quad
  y_{1,2}=\fr{\omega_{1,2}}{E}= \fr{p_T}{2E} \left(
  \mathrm{e}^{\pm\eta_3}+\mathrm{e}^{\pm\eta_4}\right)\,.
\ee

With the standard transformation
\begin{displaymath}
  \fr{\partial}{\partial y_1\partial y_2 \partial \cos\theta}
  \equiv\fr{\varepsilon_3 \varepsilon_4 \partial}{p_T\partial
  p_{L3}\partial p_{L4}\partial p_T}
\end{displaymath}
the integrand of Eq.\ (\ref{prottot}) (after integrating over
$Q_i^2$) can be considered as the distribution over the momenta
of produced photons.  Then the differential cross section of
$\ggam$ production via monopole loop can be presented in the
form
\bea
  \varepsilon_3\varepsilon_4
  \fr{d^6\sigma}{d^3p_3 d^3p_4}&=& \left(\fr{\alpha}{\pi}\right)^2
  y_1^2f(y_1)y_2^2f(y_2)
  \fr{5RE^4}{112\pi}\delta^{(2)}({\bf{p}}_{T3}
  +{\bf{p}}_{T4})\Phi\,,  \\
  \Phi &=&\left(4-\fr{p_T^2}{\omega_1\omega_2}\right)^2\equiv
  \left(4- \fr{1}
  {\cosh^2\left[\left(\eta_3-\eta_4\right)/2\right]}\right)^2 \,.
  \nonumber
\eea
Integrating over one transverse momentum and azimuthal angle
the cross section is given by
\be
  \fr{d^3\sigma}{d\eta_3d\eta_4dp_T^2} = \left(\fr{\alpha}{\pi}\right)^2
  y_1^2f(y_1)y_2^2 f(y_2)\fr{5RE^4}{112}\Phi\,.
\ee

{\em $\ggam$ total transverse momentum distribution.}
The total transverse momentum of the produced photon pair ${\bf{k}}_T
\equiv {\bf{p}}_{T3}+{\bf{p}}_{T4}$ is equal to the sum of
transverse momenta of virtual photons, ${\bf{k}}_T={\bf{q}}_{T1}+
{\bf{q}}_{T2}$. The latter are related to the photon virtualities by
\be
  {\bf{q}}_{Ti}^2=(1-y_i)(Q^2_i-Q^2_{i\,\mathrm{min}})\,, \quad i=1,2
\ee
where $Q^2_{i\,\mathrm{min}}$ is the minimal value of $Q^2_i$ for
energy fraction $y_i$ as given in (\ref{min}).

Since characteristically $Q_i^2\ll \hat s$, the photon pair
transverse momentum is typically much smaller than the
transverse momenta of the produced photons: $k_T\ll
p_{T3}\approx p_{T4}$. Therefore, the distribution in $p_{Li}$,
$p_T$ factorizes from that in $k_T$. The latter distribution can
be obtained by integration over the virtual photon fluxes only
(changing the order of integration). The detailed form of this
dependence would be an additional test for the origin of the
discussed photons. The corresponding calculations are simple but
cumbersome, and one can postpone them to the time
when first events of discussed type are observed.

However, even 
before performing the calculation we can conclude that this distribution
is peaked near $k_T=0$. This is evident for the elastic
contribution where the scale of the distribution is limited from
above by that of the form factor. For the inelastic contribution
the virtual photon flux distribution is wider, however the mean
value of $k_T^2$ is much lower than $\hat{s}$.

\section*{\em Acknowledgments}

We acknowledge fruitful discussions with Gregory Landsberg and
Valery Serbo. We thank Rainer Scharf for a critical reading of
the manuscript. I.F.G. is grateful to David Finley, Peter Lucas
and Hugh Montgomery for their warm hospitality while staying 
at FNAL.  Besides, {\em this work is supported by grants
of INTAS -- 93 -- 1180ext, RFBR 96-02-19079 and Volkswagen
Stiftung I\hspace{0.5mm}/72 302}.

\newpage
\end{document}